\documentclass{article}
\usepackage{authblk, amsmath, amsthm, todonotes, booktabs, float, hyperref, comment}

\begin{document}
\title{Modeling Oyster Reef Reproductive Sustainability: Analyzing Gamete Viability, Hydrodynamics, and Reef Structure to Facilitate Restoration of \textit{Crassostrea virginica}}
\author[1]{Justin Weissberg\thanks{The authors would like to thank Dr. James Browne and Cody Onufrock for their expertise and assistance.}}
\author[2]{Vinny Pagano}
%\author[3]{Dr. James P. Browne}
\affil[1]{Brandeis University}
\affil[2]{Princeton University}
%\affil[3]{Department of Conservation and Waterways, Town of Hempstead [this order might be reversed]}
\date{January 2021}
\maketitle

\begin{abstract}
The eastern oyster is a keystone species and ecosystem engineer. However, restoration efforts of wild oysters are often unsuccessful, in that they do not produce a robust population of oysters that are able to successfully reproduce. Furthermore, the dynamics of wild oyster fertilization is not yet well understood. Through conducting an experiment predicated on quantifying the influence of elementary aspects of fertilization kinetics—sperm concentration, gamete age, and success rate—we found that, as stochastic as the mating process may seem, there are correlations which fundamentally serve as the framework for assessing long-term sustainability, reef structure, and hydrodynamic parameters in relation to fertilization. We then focused on mathematically defining a procedure which simulated a concentration distribution of a single sperm and egg release where there existed conditions necessary for breeding to take place. We found a very significant impact of both gamete age and sperm concentration on fertilization rate (\(p<0.0001\)). Our hydrodynamic model demonstrates that distance can also drastically influence broadcast spawning. This could be used as a foundation for developing a flexible model for wild oyster fertilization based on placement, initial seawater conditions, and size of the starting population. The results of this research could be implemented into a more user-friendly program which would accept multiple variables as inputs and output the probability of fertilization given arbitrary values. By accounting for environmental deviations, this generalization would increase its compatibility with the public and actualize the project’s intended purpose: enhance the planning of oyster reef restoration projects. 
\end{abstract}

\tableofcontents

\section{Introduction}
The eastern oyster (\textit{Crassostrea virginica}) acts as an ecosystem engineer and is a keystone species (Sanjeeva, 2008). Therefore, the eastern oyster is a source of invaluable natural capital. Humans rely on earth’s natural capital to carry out vital ecosystem services, including water filtration, storm surge protection, and habitat creation. The oyster provides habitats and nutrition for a variety of organisms, and studies have shown that established reefs enhance the species’ richness (Luckenbach et al., 2005). Anecdotally, organisms such as crabs, amphipods, and some fish rely on oyster reefs for survival (Grabowski, 2004). Oyster reefs increase habitat complexity; many organisms depend on their shell structure for shelter and refuge from predators (Coen and Luckenbach, 2000; Rodney and Paynter, 2003). Additionally, oysters help to filter excessive amounts of phytoplankton present in the water (Lenihan et al., 1996; Jackson et al., 2001). They also play a valuable role in nitrogen cycling in estuaries. By absorbing excess nitrogen in the water, they help to reduce instances of eutrophication brought about by human nitrogen pollution (Newell et al., 2005, Sebastiano, 2013). Furthermore, oyster reefs may also act as barriers to erosion via absorbing wave and storm surge energy (Gereffi et al., 2012). They are important, hard barriers which protect Spartina marshes from erosive forces (Meyer et al., 2008). Over long periods of time, oysters sequester carbon from fossil fuel emissions in the form of calcium carbonate (Grabowski and Peterson, 2007). 

Although the eastern oyster was once numerous and prolific, this species is now struggling in most major bodies of water (Beck et al., 2011). It is estimated that 85\% of all-natural reefs around the world have been destroyed (Grabowski and Peterson, 2007; Beck et al., 2011). Consequently, humans have invested millions of dollars into restoration efforts, most notably in the Chesapeake Bay, Gulf of Mexico, Delaware Bay, and Hudson River (Hargis and Haven, 1988; Frankenberg, 1995, Kreeger, 2011). There are a myriad of  ongoing restoration projects, particularly in the New York waters, where oysters are both a profitable commercial enterprise and a natural bio filter (Stark et al., 2011).

The conditions which have led to the worldwide ecological decline of wild oysters can be either natural or anthropogenic. Oysters have struggled due to excessive harvesting, destruction of habitats, and two main diseases: \textit{Hasplosporidium nelsoni} (MSX) and \textit{Perkinsus marinus} (Dermo) (Bosch and Shabman, 1990; Horton and Eichbaum, 1991; MacKenzie, 1996; Paynter, 1996). Humans appear to be the primary cause for the decline of oyster reefs because they have caused the reefs to become too sparsely populated, thus making sufficient external fertilization virtually impossible (Lenihan and Peterson, 1998; Lenihan and Peterson, 2004). As a prominent commercial product, oysters have been overharvested for food and their calcium carbonate shells, which have various commercial applications. The reduction of oyster populations, insofar as too few adult oysters are available to sustain a reef, has become nearly pervasive. Therefore, an inadequate number of larvae is produced, thereby jeopardizing reef sustainability (Mann and Powell, 2007). This paradigm is then magnified by a poor habitat for recruitment due to human interaction. Naturally, oysters grow in vertical reefs, yet humans break up, spread out, and even dredge these reefs for the attainment of oysters of ideally marketable size and shape (Kennedy, 1995; Gosling, 2003).

\begin{comment}
Our results may indicate that harvesting of oysters disrupts the reef habitat and decreases the oyster population until there are too few for the reefs to sustain themselves. 
\end{comment}

Due to an insufficient understanding of the influence of gamete quality on fertilization, research pertaining to the reproduction of oyster populations has proven to be highly variable (Boudry et al., 2002). Although there have been extensive resources invested into the reproduction process of fresh and marine organisms, little research has been performed on mollusks and especially bivalves. Likewise, restoration efforts have been hampered due to an unsatisfactory understanding of the factors that affect fertilization in the wild. Our findings are novel in that they demystify oyster fertilization, illuminating the scholarly conversation and presenting an opportunity to restore a once prevalent species.

In order to extract information regarding fertilization success in free-spawning organisms such as the eastern oyster, two approaches can be used (Denny and Shibata, 1991): (1) field data can be collected during a spawning event, or (2) models can be constructed based on both laboratory experiments about fertilization and field observations about water flow conditions. Issues related to an in situ approach includes experimental artifacts that may skew estimates of fertilization (Pennington, 1985; Yund, 1990; Levitan, 1991; Levitan et al., 1991, in press; and Denny and Shibata, 1991). Secondly, that natural spawning events are rare makes it especially difficult to collect adequate data (Petersen et al., 2008). Limitations on constructing fertilization models include the difficulty in collecting accurate flow data and the knowledge of how gametes behave in spawn-like conditions. In the laboratory, we investigated the interaction between gamete age and sperm concentration on overall zygote production. There has been an abundance of literature directed at describing how gamete number, age, and other factors impact fertilization in invertebrates (Lilie, 1915; Rothschild and Swann, 1951; Hulton and Hagstrom, 1956; Brown and Knouse, 1973; Vogel et al., 1982; Pennington, 1985). However, in regard to \textit{Crassostrea virginica}, literature on this topic is limited.

We collected data on the interaction of gamete age and sperm concentration in the lab to accurately model oyster fertilization in the wild. Since zygotes form though external reproduction via broadcast spawning, the egg and sperm originate from different oysters, making the process random and susceptible to confounding variables such as swimming speeds and anisotropic conditions, notably seawater turbulence. Establishing a robust model for predicting a reef’s overall success rate in a controlled environment may minimize other environmental factors from being overlooked. This is a first step toward eliminating the problems which currently revolve around wild oyster reintroduction.

\subsection{The Importance of Structure in Biological System}
The physical structure of an environment is known to have profound implications on population biology and interactions between organisms (Bell et al., 1991). The spread and abundance of a species throughout a habitat is directly influenced by structure, due to the availability of substrate for colonization or recruitment (Underwood and Denley, 1984). Furthermore, habitat structure can provide protection from competition (Huffaker, 1958). The physical habitat structure has major biological implications since it plays a relevant role in the coupling of physical and biological variables  (Belsky et al., 1989; Mann and Lazier, 1991). The coupling of physical and biological habitat variables has many ecological consequences that are greatly appreciated, especially in the marine environment. Studies have analyzed how habitats affect flow regimes (Vogel, 1981), salinity (Klinne, 1963), and dissolved oxygen concentration (Diaz and Rosenberg, 1985). In terms of oyster reefs, the marine benthic habitat contributes to the settlement of larvae (Breitburg et al., 1995), recruitment (Eckman, 1993), filtering of food (Muschenheim, 1987; Butman et al., 1994; Sanford et al., 1994), predation (Skilleter and Peterson, 1994), and the composition of a community (Wildish and Kristmanson, 1979; Baynes and Szmant, 1989; Leonard et al., 1998). Physical habitat structure has been shown to affect physical variables such as flow velocity, which has been shown to have a significant influence on the performance of the species that resides in the habitat (Lenihan, 1999). An improved understanding of the role of physical habitat structure on the reproduction of organisms such as the eastern oyster will allow for an increased ability to manage and replenish destroyed reefs using ecological engineering (Jones et al., 1994). While successful habitat restoration can potentially be achieved by recognizing and rebuilding the functional properties of a habitat, there is still limited literature explaining how habitats function (Thayer, 1992).
The purpose of this research is to set forth the foundation for determining the critical mass and reef structure necessary to maintain the long-term integrity of a reef. This exploration targets the influence of gamete age and sperm dilution on fertilization success. It is likely that, as sessile organisms, oysters depend on gamete age, dilution, hydrodynamics, and distance between reefs for successful fertilization (Denny, 1988; Denny and Shibata, 1989). 

\section{Methods}
\subsection{Conditioning and Gamete Extraction}
On June 24, 2014, twenty-five eastern oysters, donated by the Town of Hempstead Department of Waterways and Conservation, were used in this spawning experiment. The researchers at the lab fed them two types of microalgae: \textit{Tetraselmis chui} and \textit{Isochysis galbana}. The oysters were conditioned for eight weeks in tanks with a constant flow of water from Reynolds Channel, as well as a gradual increase in water temperature up to approximately \(18^\circ\)C to ensure gamete maturation (Davis and Loosandoff, 1952).
The dry stripping method was used to extract the gametes (Allen and Bushek, 1992). In order to extract the gametes, each oyster was shucked open. A pipette was used to puncture the gonad and subsequently extract the gametes. A light microscope was used in order to identify the type of gamete (egg or sperm). The sperm were pipetted into one stock solution and eggs were pipetted into another. This method helped to optimize the precision of dilutions and aging of the gametes.

\subsection{Creating Different Gamete Ages and Sperm Dilutions}
Once the gametes were separated into two stock solutions, different dilutions of sperm were created. Dilutions of 1:10 were used in order to produce four dilutions (ten percent, one percent, one tenth of one percent, and one hundredth of one percent). The gametes were also mixed at varying intervals in order to manipulate gamete contact time. These include 25, 45, 65, 125, 185, and 245 minutes. Each sample received the same amount of egg (0.5 ml) at a concentration of twelve percent. Each sample also received 0.2 ml of sperm with the sperm concentration varying between the aforementioned dilutions. Lastly, 5 ml of filtered seawater was added to prevent clumping of gametes. The samples were transferred from plastic cups to glass containers for storage in a refrigerator at \(2-4\)$^\circ$C. Lugol’s iodine was added to every sample for the purpose of preservation.

\subsection{Data Collection}
A microscope and a Sedgewick-Rafter counting cell were used to determine the total quantity of eggs, fertilized eggs, and unfertilized eggs for each sample. The average egg concentration was also obtained.

\begin{figure}[htp]
    \centering
    \includegraphics[width=4cm]{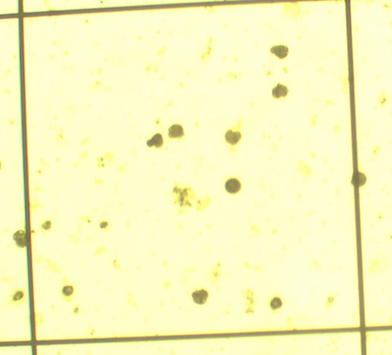}
    \caption{A sample containing gametes that were aged for 25 minutes and had a sperm dilution of \(0.1\).}
    \label{fig:counting}
\end{figure}

\subsection{Statistical Analyses}
In order to determine the curve of best fit for the data, exponential regressions were utilized, and \(R^2\) values were calculated to analyze the strength of the correlations. A multi-factor analysis of variance (ANOVA) was performed to determine the significance of the dilution and gamete age on fertilization. Lastly, a forward step-wise regression was used to create a model for extrapolating the fertilization rate of eggs.

\subsection{Model Design}

Subsequent to our fertilization trials, we began refining a mathematical model in 2018. The model can be divided into two components. First, the empirical fertilization data was acquired, which allowed for the quantification of the influence of gamete age (time after gametic release) and sperm concentration. Next, in an effort to parametrize gamete survival in a fluid dynamics model, the data was analyzed using a forward step-wise regression in both gamete age and concentration variables. This regression model was necessary to estimate fertilization in the wild, as it was applied to Csanady's Equation to produce a fertilization model in a hydrodynamic context (Csanady, 1973).
\\\\\\
\subsection{Outline of Simulation}
Consider one of the simplest cases in which fertilization can occur:
\\

\begin{minipage}{.65\textwidth}
\begin{itemize}
\item Two oysters are present within the model
\begin{itemize}
    \item One oyster releases sperm and the other releases eggs 
    \item The total mass of each release is denoted by \(M\)
    \item The difference between swimming speeds of gametes is taken as trivial
\end{itemize} 
\item	Oysters rest on the same leveled platform (two-dimensional model)
\item	The gametic releases from the oysters are modeled as diffusion equations, with each function being relative to a coordinate axis
\item	There are isotropic conditions (steady-state), meaning turbulence is nonexistent and fluid dynamics will follow basic seawater principles
\item	The centers of both graphs of the gametic releases are a distance \(r\) apart
\begin{itemize}
\item	In the egg equation, a variable can be appended to the spatial variables as a functional phase shift
\end{itemize}
\end{itemize}
\end{minipage}
\begin{minipage}{.25\textwidth}
\begin{figure}[H]
\includegraphics[scale=0.6]{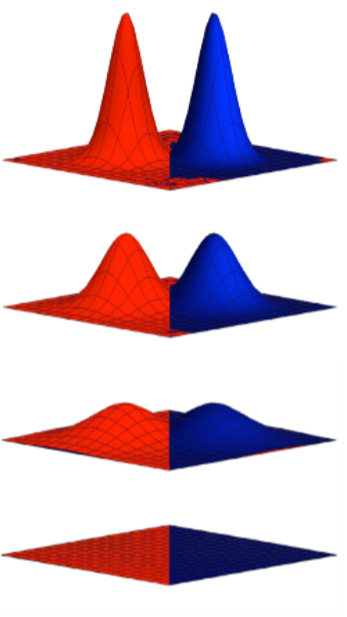}
\caption{Time evolution of elementary diffusion model.}
\label{fig:evolution}
\end{figure}
\end{minipage}
\\\\
These conditions can easily be generalized to simulate diffusion for three spatial variables. However, to model wild oyster fertilization, it suffices to use only one diffusion equation for the sperm gametes:
\begin{equation}\label{eq 1}
c(r,t) = \frac{M}{\sqrt{4 \pi D t}} \exp\Big({-\frac{r^2}{4 D t}}\Big).
\end{equation}
In other words, given a quantity \(M\) of a substance, we are able to compute its concentration \(c(r,t)\) at time \(t\) and distance \(r\) from the point of release.

Naturally, more precise models may involve both classes of gametes to capture more nuanced behavior. For instance, the three-dimensional analogue of the previous equation commonly used for anisotropic models is the following:
\begin{equation}\label{eq 2}
c(x,y,z,t) = \frac{M}{(\sqrt{4 \pi D t})^3} \exp{\Big(-\frac{x^2 + y^2 + z^2}{4 D t}\Big)}.
\end{equation}
It is easy to verify that we satisfy the law of conservation of mass since
\begin{equation}
\int_{-\infty}^\infty c(r,t) \hspace{.5 em} dr = \int_{-\infty}^{\infty}\int_{-\infty}^{\infty}\int_{-\infty}^{\infty} c(x,y,z,t) \hspace{.5em} dzdydz = M.
\end{equation}
More specifically, if \(\nabla^2\) denotes the Laplace operator and we consider an anisotropic medium such as the ocean, we can obtain a solution corresponding to an instantaneous (\(t = 0\)) and localized (\(x = y = z = 0\)) release:
\begin{equation}\label{hip}\frac{\partial c}{\partial t} = - \vec{\nabla} \cdot \vec{q} = \vec{\nabla} \cdot (D \vec{\nabla} c) = D \hspace{.3 em} \nabla^2 c\end{equation}
\begin{equation}c(x,y,z,t) = \frac{M}{(\sqrt{4 \pi D t})^3 \sqrt{D_x D_y D_z}} \exp\Big(-\frac{x^2}{4D_x t} - \frac{y^2}{4D_y t} - \frac{z^2}{4D_z t}\Big),\end{equation}
where \(\vec{q} = \langle q_x,q_y,q_z\rangle\) and \(D\) are the flux and the diffusion coefficient of the substance, respectively (Diffusion, 2012).

Note that time can have arbitrary units and will
affect the diffusion coefficient accordingly in the solutions to the partial differential equations. Moreover, the process of fertilization will be modeled using the first diffusion equation.
\begin{figure}[htp]
    \centering
    \includegraphics[width=9cm]{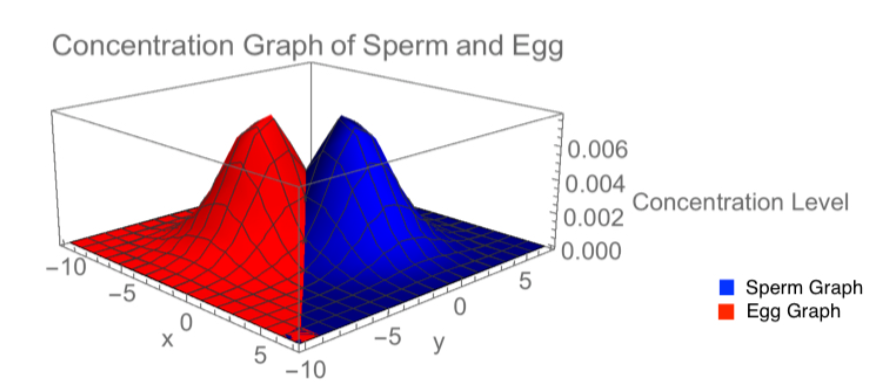}
    \caption{Elementary diffusion model at fixed time.}
    \label{fig:demo}
\end{figure}

\begin{comment}
Diffusion Constant (Stokes-Einstein Equation)
The diffusion of spherical particles through a liquid with a low Reynolds number takes the form
where  is the Boltzmann constant, is the temperature, is the dynamic viscosity (solvent viscosity), and is the radius of the spherical particle, which is to be assumed. This models the theoretical and chaotic Brownian motion of particles in a liquid medium. Molecular (dynamic) viscosity is analogous in laminar flow to eddy viscosity in turbulent flow because when both types are present, the extreme difference, which is in orders of magnitude, results in negligibility.

\end{comment}

In a marine environment, sperm and ova will be mixed due to turbulence. However, they will significantly dilute over time. The concentration of sperm and egg in relation to oceanic turbulence can be estimated. Egg concentration has been shown to have an insignificant effect on fertilization (Levitan et al., 1991) and, therefore, egg concentration is not included as a variable in the fertilization regression model.

Our mathematical contribution expounds upon Equation \ref{eq 2} in a more realistic setting. Namely, we adopt an experimental perspective and modify the seminal research of Denny and Shibata, applying their methodology to \textit{Crassostrea virginica}. As defined in Equation \ref{Fert}, we consider \begin{equation}
F(S(x,y,z),t),
\end{equation} where \(F\) denotes the probability of fertilization in space-time and \(S(x,y,z)\) represents the sperm concentration as defined by Csanady’s model (Csanady, 1973; Lauzon-Guay and Scheibling, 2007): 

\begin{equation}\label{Denny}
S(x,y,z) = \frac{Q_s \bar{u}}{2\pi \alpha_y \alpha_z u_*^2 x^2}\Big(e^{- \frac{y^2 \bar{u}^2 \alpha_z^2 + (z-s)^2\bar{u}^2 \alpha_y^2}{2\alpha_y^2 \alpha_z^2 u_*^2 x^2}} + e^{- \frac{y^2 \bar{u}^2 \alpha_z^2 + (z+s)^2\bar{u}^2 \alpha_y^2}{2\alpha_y^2 \alpha_z^2 u_*^2 x^2}}\Big).
\end{equation}
Notably, our simulation is an improvement of Equation \ref{Denny}, both for the purposes of our regression and for future models which build upon correlations such as gamete concentration, because of its efficient multidimensional averaging implementation. By excluding unanticipated, emergent outliers arising from the choice of variables in Table \ref{tab:values for sim}, we also allow for a more accurate simulation that is generalizable to other aquatic species.
\begin{table}
    \centering
    \begin{tabular}{l|c|c|c}
        Variable Name & Symbol & Value \\
        \toprule
        Horizontal Diffusion Coefficient & \(\alpha_y\) & 2.25\\
        Vertical Diffusion Coefficient & \(\alpha_z\) & 1.25\\
        Sperm Release Rate & \(Q_s\) & \(2.33 \times 10^6\) \\
        Current Velocity & \(\bar{u}\) & 0.5\\
        Frictional Velocity & \(u_*\) & 0.05\\
        Height of Gamete Release & \(s\) & 0.3\\
        Positional Concentration of Sperm & \(S\) & \(-\)\\
    \end{tabular}
    \caption{Values for simulation.}
    \label{tab:values for sim}
\end{table}

\subsection{Turbulent Flow in Advective Environments}
In applicative hydrodynamics, oceanic eddy turbulence is caused when a wave traverses a barrier that is not parallel to its velocity vector. It can create an environment that is temporarily more prone to mixing. The conditions under which an orbital vortex may take place are not well-defined, making any mathematical model of oyster diffusion somewhat unfinished or inconclusive. However, simplifications can confine the reaction to a set number of reacting elements to predict the overall behavior of the chaotic system.

Consider an anisotropic system with an intrinsic time scale. The time it takes for a particle to cover a circular orbit of circumference \(\pi d_o\) with a nominal orbital velocity \(u^*\) is:
\begin{equation}\tau = \frac{\pi d_o}{u^*} \sim \frac{d_o}{u^*}.\end{equation}                                                
Furthermore, the relationship in a random-walk model for the eddy diffusion coefficient is:
\begin{equation}\varepsilon \sim d_o u^*.\end{equation}                                         
The physics of these equations, of course, only hold for perfectly circular orbits.
Let \(\varepsilon_i\) and \(D\) represent axis turbulent transport and molecular transport in a standard mass advection-diffusion equation respectively. (Note that mass can be synonymous with heat transfer.) Once expanded to three dimensions and since \(\varepsilon \gg D\), we have from Equation \ref{hip} that
\begin{equation}\frac{\partial c}{\partial t} + \vec{u} \cdot \vec{\nabla} c = \varepsilon \nabla^2 c\end{equation}
\begin{equation}\frac{\partial c}{\partial t} + u\frac{\partial c}{\partial x} + v \frac{\partial c}{\partial y} + w \frac{\partial c}{\partial z} = \frac{\partial}{\partial x}\Big(\varepsilon_x \frac{\partial c}{\partial x}\Big) + \frac{\partial}{\partial y}\Big(\varepsilon_y \frac{\partial c}{\partial y}\Big) + \frac{\partial}{\partial z} \Big(\varepsilon_z \frac{\partial c}{\partial z}\Big),\end{equation} 
where \(u\), \(v\), and \(w\) are the corresponding components of the velocity vector \(\vec{u}\) for an arbitrary current. The representation of the eddy diffusion coefficient is currently abstract because of its natural randomness and variation (Roberts and Webster, 2003). It follows that the general solution corresponding to an instantaneous and localized release is:
\[c(x,y,z,t) = \frac{M}{(\sqrt{4 \pi D t})^3} \exp \Big(- \frac{(x-ut)^2 + (y-vt)^2 + (z-wt)^2}{4Dt}\Big).\]
This concludes all of the mathematically defined relationships for diffusion.
\section{Results}
\begin{table}[ht]
\centering
\begin{tabular}[t]{c | c c c c}
\multicolumn{1}{c}{Gamete Age (minutes)} &
\multicolumn{4}{r}{Sperm Concentration (decimal)}\\
\toprule
& 0.1000 & 0.0100 & 0.0010 & 0.0001\\
\midrule
25 &0.470&	0.270&	0.280&	0.179\\
\hline
45&	0.392&	0.235&	0.223&	0.147\\
\hline
65&	0.328&	0.204&	0.188&	0.121\\
\hline
85&	0.274&	0.178&	0.154&	0.099\\
\hline
145&	0.159&	0.117&	0.0844&	0.054\\
\hline
205&	0.093&	0.077&	0.046&	0.029\\
\bottomrule
\end{tabular}
\caption{Mean fertilization rate for sperm concentrations and gamete ages.}
\label{tab:main data}
\end{table}
\subsection{Sperm Dilution and Gamete Age}

The highest fertilization was recorded for the 10\% sperm dilution. A significant decrease in fertilization occurred due to sperm dilutions (\(p = 2.03 \times 10^{-16}\)). The lowest fertilization was achieved when the gametes were aged to 205 minutes. Gamete age alone appeared to have a significant effect on fertilization rate (\(p = 1.26 \times 10^{-17}\)). Lastly, results showed a significant interaction between gamete age and sperm dilution (\(p=0.0071\)).

\subsection{Fertilization Regression}
Using these fertilization results from Table \ref{tab:main data} in a regression, it was found that both sperm concentration and gamete age variables were highly significant ($p < 0.0001$). The regression created from the data is: \begin{equation} \label{Fert} F = 1.71153 \cdot S - 0.00149 \cdot A + 0.31334,
\end{equation}
where $F,$ $S,$ and $A$ represent the probability of fertilization, sperm concentration (as a decimal proportion of the starting concentration), and gamete age in minutes, respectively. This regression model provides an estimate fertilization under different temporal and concentration parameters for use in the hydrodynamic model. This method differs from others such as Denny and Shibata who modeled the process of fertilization and suggested only about one percent of an individual egg's surface is fertilizable (Denny and Shibata, 1989). 
%\subsection{Numerical Confirmation}
%We used Mathematica to evaluate the upper bound time intervals. All results yielded as the maximum value, which is the general limit for  for negligibly low sperm concentrations. We also utilized Mathematica to compute the integration solutions to the partial differential diffusion equations, which we predicted to decrease: Table 2: A table of the solutions to the analytical method involving the diffusion equations in three-dimensional space. The distance  will be acquired by multiplying every value of by .
\begin{figure}[htp]
    \centering
    \includegraphics[scale = .4]{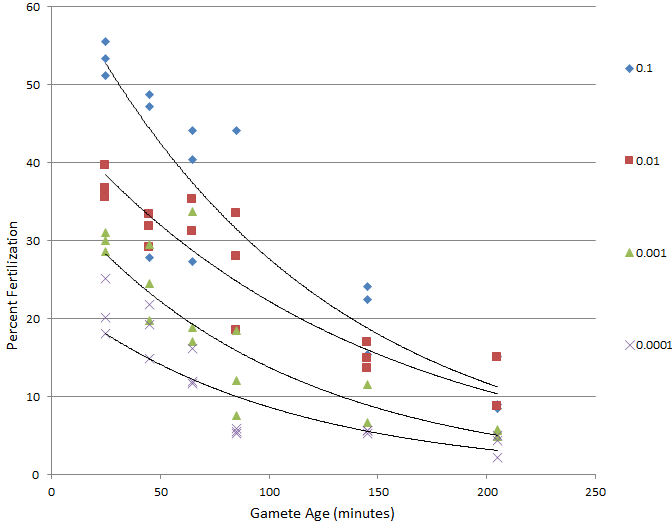}
    \caption{Percent fertilization at varying sperm concentrations and gamete ages with exponential regression trendlines.}
    \label{fig:nitty gritty}
\end{figure}
\begin{table}[H]
    \centering
    \begin{tabular}{c|c|c|c}
    Concentration & Fitted Equation & \(R\) & \(R^2\)\\
    \toprule
         \(0.1000\) &  \(65.348 \cdot e^{-0.009 \cdot x}\) & 0.9266 & 0.8586\\
         \hline
         \(0.0100\) & 4\(46.125 \cdot e^{-0.007 \cdot x}\) & 0.9373 & 0.8786\\
         \hline
         \(0.0010\) & \(35.961 \cdot e^{-0.010 \cdot x}\) & 0.9105 & 0.8290\\
         \hline
         \(0.0001\) & \(23.070 \cdot e^{-0.010 \cdot x}\) & 0.8800 & 0.7747\\
         \bottomrule
    \end{tabular}
    \caption{Linear regression statistics corresponding to Table \ref{tab:main data}}
    \label{tab:regression}
\end{table}

\begin{figure}[htp]
    \centering
    \includegraphics[scale=.5]{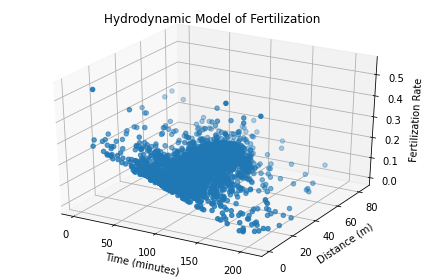}
    \caption{Fertilization hydrodynamics model produced in Python.}
    \label{fig:model}
\end{figure}

\section{Discussion}
To the authors' knowledge, this model is the first of its kind to use empirical data as a baseline for modeling external fertilization in bivalves. The empirical data indicated higher fertilization rates compared to fertilization models of sea urchins (Denny and Shibata, 1989). Unlike sea urchins, oysters are completely sessile and cannot move toward other oysters to increases chances of successful fertilization. Therefore, oysters must rely on the quantity of gametes they produce to facilitate fertilization.

There are a multitude of other factors which could be mathematically implemented into the procedure to make it more specific. The fertilization equation has been shown to essentially be time-dependent (in many cases, the sperm concentration is negligible when finding maximum positive time). Turbulence generally increases the diffusion coefficient, which could allow the partial differential equations to yield higher probabilities of fertilization. Furthermore, oysters tend to switch sex halfway through their lifespan, and therefore the long-term modeling is made significantly difficult. Molecular viscosity appeared to be an impractical method to use for the diffusivity, but this conclusion may be subject to change as the precision of analysis increases.

%As a start, if the expected percentage of zygotes formed for a distance is under 50\%, the dynamical system in isotropy could be completely unsustainable. Results show that distances over 8µ correspond to unsustainable environments. 

\subsection{Gamete Swimming Speed}

\begin{figure}[H]
\centering
\includegraphics[scale = .6]{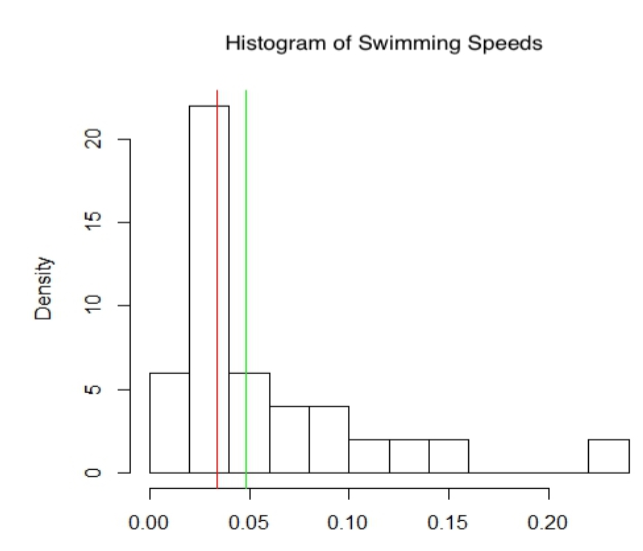}
\caption{Adjusted histogram of sperm swimming speeds (Mann and Luckenbach, 2013).}
\label{fig:Histogram}
\end{figure}

The work of Mann and Luckenbach suggests that the sperm swims about a trillion times faster than the egg (Mann and Luckenbach, 2013). However, the validity of this experiment could be brought into question. Mann and Luckenbach only used 25 of the 30 samples---namely, the ones for which the sperm travelled a measurable distance. Even when disregarding the small sample size of the resultant data, it seems unreasonable to reject those 5 samples from mathematical models involving sperm swimming speeds because of how delayed or circular movement is an acceptable feature of Brownian motion. As such, removing the samples would create a pseudo-distribution of values for the sperm swimming speeds that is not representative of the measured data. Hence, we added the 5 events of 0.0 movement back (See Figure \ref{fig:Histogram}). 

This new histogram is a skewed-right distribution and reveals a high outlier in the data that was not accounted for in the original graph. Using only the 25 data points, the mean of the estimates is 0.058 mm/sec, which lies within the value of \(0.057 \pm 0.01\) mm/sec that Mann and Luckenbach reported. However, if the non-motile sperm are also included, then a more realistic mean of 0.048 mm/sec and median of 0.034 mm/sec are found. Regardless, further analysis and experimentation can be conducted to ensure a more accurate and reliable average for sperm swimming speeds.

\begin{comment}
Given that the temperature for oyster spawning is \(20 - 30^\circ\)C, a value of 24$^\circ$C will be used, which is consistent with the work of Mann-Luckenbach on sperm swimming speeds (Mann and Luckenbach, 2013). Mathematica will be utilized to simulate the equations for wild oyster fertilization, starting with the isotropic conditions in Laminar Flow. Oyster sperm are described in Eble and Scro (1996) as having a head of approximately \(2-3\) µm and a flagella tail of up to 40 µm in length. We will assume that the particles are spherical and have a diameter of 2.5 µm. The egg diameter is approximately 50 µm.
\end{comment}

\subsection{Facilitating Restoration}
The results and preliminary hydrodynamic models are particularly promising, as they have direct real-world implications on oyster reef restoration. Prior to this study, there was very little information available for researchers looking to most successfully engineer self-sustainable oyster reefs. Having now established several novel correlations pertinent to fertilization, we present the reader with a clearer picture.
	
The distance between oysters is paramount to permitting interbreeding within sub-populations of a reef. Moreover, in situ experiments have revealed that increasing the distance between spawning organisms of opposite sex also results in an exponential decline in fertilization (Denny and Shibata 1989; Pennington 1985; Yund 1990; Levitan 1991; Marshall 2002; Metaxas 2002). In these experiments, just a few meters of separation between spawning organisms resulted in a significant reduction in fertilization success. This study supports past work showing that fertilization is considerably dependent on the distance between organisms of the opposite gender. For example, at a distance of 80 meters and gamete age of at least 85 minutes, our model yields an average fertilization value of only 6.38\%. Thus, if a researcher constructs sub-populations that are 80 meters apart, they will likely be unable to interbreed with each other---which may eventually lead to the collapse of the entire reef and potentially the ecosystem.
\begin{table}[t]
\centering
\begin{tabular}{c|c}
     Length (mm)& \(p_f\)\\
     \toprule
     25 & 0.100\\
     \hline
     35 & 0.205\\
     \hline
     45 & 0.246\\
     \hline
     55 & 0.400\\
     \hline
     65 & 0.550\\
     \hline
     75 & 0.562\\
     \hline
     85 & 0.583\\
     \bottomrule
\end{tabular}
\caption{Data from Powell et al. (2013) showing the combined proportion of females (\(p_f\)) across several sub-populations by size.}
\label{tab:Powell data}
\end{table}

Although this research did not address different gender ratios, this may also be a non-trivial factor to consider when constructing reefs. Management efforts must make sure that reefs are constructed with a sufficient ratio of males to females. It is interesting to note that, while the eastern oyster is capable of beginning as a male and transitioning to a female throughout their life cycle, they can even switch sex in response to the environment (Buroker, 1983; Kennedy, 1983). Based on combined measurements from Delaware Bay that were published by Powel et al. (2013), oysters of 25 mm average only 10\% female, but increase to over 70\% past the 90 mm size (Table \ref{tab:Powell data}). When combined with our model results, this implies that a constant addition of new recruitment is needed to maintain a high ratio of small young individuals that will represent a higher male to female ratio. The sparser the population, the more critical it becomes that a high ratio of new recruits is maintained. Restorations therefore must be supplied with a growing percentage of new individuals for several years in order to succeed.

\subsection{Model Limitations}
Tidal estuaries are dynamic environments in which water velocities are contingent upon local bathymetry, sun and moon orbital phases, and meteorological activity. Thus the velocity differs by time and location. Directions of current flow can also vary depending on tidal cycles. Rates of spawning in the wild are highly variable and seldom predictable. In a mass spawning event, most oysters may spawn but not necessarily the entire population. Other relevant environmental parameters such as predation, temperature, and salinity should be incorporated into the model. Settlement of larvae can also be used to estimate the recruitment rate of an oyster population. The model should be tested against an actual spawning episode in the wild, and sub-samples should be collected to evaluate its robustness.
\section{Conclusion}
	This model is based on gamete kinetics and should be compared to field estimations of actual fertilization success. Observations regarding natural spawning are also required in order to effectively model fertilization in the wild (Pennington, 1985; Yund, 1990; Grosberg, 1991; Levitan, 1991). This model helps explain how human overharvesting could have led to the decline of natural oyster reefs. As shown by this model, the reefs are very sensitive to the distance between the individual oysters. A few meters appear to be critical for fertilization and thus the failure or success of a reef. Humans should learn and adapt from their mistakes by using the same principles of population density to help restore oyster beds. By applying the results of this investigation to restoration efforts, reefs can be more effectively and robustly engineered. Humans have benefited greatly commercially and environmentally from the oyster. Now there is an opportunity where humans can give back by helping to restore this keystone species.             

\begin{comment}
We have successfully constructed a mathematical model for wild oyster fertilization which can be generalized to k-dimensional space. Scientists can now utilize this equation/procedure as preliminary research to accurately determine the sustainability and sufficiency of their oyster reefs. We used the solutions of two partial differential equations to simulate an isotropic environment to find the approximate multivariable intersections and analyze their trend in the context of oyster fertilization. However, higher computation power will be required to compute the surface integrals for when the diffusion constant is represented by the Stokes-Einstein Equation. We found that the system is potentially sustainable at a distance that is around . These results are reflective of how sensitive the random-walk model is. 
The importance of distance in wild oyster fertilization cannot be further stressed. Reintroduction of wild oysters are dependent on several factors, many of which may have led to a steep decline in the oyster population over the last century.
\end{comment}

\subsection{Future Research}

Conducting experiments on sperm swimming speeds and further testing Equation \ref{Fert} at different temperatures (\(20-30^\circ\)C) will increase the precision, accuracy, and scope of our results. One compelling implementation of a robust model would be toward the development of a user-friendly software designed to assist biologists by mathematically informing them about the practicality and effectiveness of their wild oyster reef setups. A Monte Carlo simulation for measuring the sustainability of oyster populations is another interesting approach, since it provides a more numerical and probabilistic interpretation of the random-walk model. For example, it could easily deal with the statistic that 80\% of wild oysters in established reefs are female and better calculate the long-term effects of population growth. Generalizing the partial differential equations to four spatial dimensions and trying to find similarities, differences, and the potential usefulness of the properties of its solutions could also allow for a sustained investigation. Incorporating more variables such as salinity of seawater, density, pressure, kinematic viscosity, swimming speeds of gametes, and eddy turbulence will make the equation more realistic and applicable. Further exploring this exponential correlation between the percentage of zygotes formed and the distance away that the sperm is from the egg should overall improve our understanding of the fertilization process. Finally, giving results that are in context and citing specific examples which can be empirically verified would further confirm the efficacy of our model.

\section{References}

\hspace{1.1em} Allee, W.C. (1931) Animal Aggregations, a Study in General Sociology. University of Chicago Press, Chicago, Illinois. 431 pp.

Allen, S.K, Bushek, D., (1992) Large scale production of triploid oysters, Crassostrea virginica (Gmelin) using “stripped” gametes. Aquaculture 103, 241-251.
	
Babcock, R., C. N. Mundy, and D. Whitehead. 1994. Sperm diffusion
models and in situ confirmation of long distance fertilization in the
free-spawning asteroid Acanthaster planci. Biol. Bull. 186: 17-28.

Bayne, T. W., and A. M. Szmant. (1989) Effect of current on the sessile benthic community structure of an artificial reef. Bulletin of Marine Science 44:545-566.

Beck, Michael W., Robert D. Brumbaugh, Laura Airoldi, Alvar Carranza, Loren D. Coen, Christine Crawford, Omar Defeo, Graham J. Edgar, Boze Hancock, Matthew C. Kay, Hunter S. Lenihan, Mark W. Luckenbach, Caitlyn L. Toropova, Guofan Zhang, and Ximing Guo. (2011). "Oyster Reefs at Risk and Recommendations for Conservation, Restoration, and Management." BioScience, 61(2): 107-116. 

Bell, S. S., E. D. McCoy, and H. R. Mushinsky,editors. (1991) Habitat structure: the physical arrangement of objects in space. Chapman and Hall, London,UK.

Belsky, A. J.,R. G. Ammundson, J. M. Duxbury, S. J. Riha, A. R. Ali, and S. M. Mwonga. (1989) The effects of trees on their physical, chemical, and biological environments in semi-arid savannah in Kenya. Journal of Applied Ecology 26:1005-1024.

Benoit Cushman-Roisin Thayer School of Engineering Dartmouth College. (n.d.). Dispersion and Mixing: Turbulent Diffusion.

Benzie, J.A.H., K.P. Black, and P. Dixon. Small-Scale Dispersion of eggs and Sperm of the Crown-of-Thorns Starfish (Acanthaster planci) in a Shallow Coral Reef Habitat. Biological Bulletin. (1994): 153-167. Web. 1 Dec. 2013.

Bosch, Darrell J., and Leonard A. Shabman. (1990) Simulation Modeling to Set Priorities for Research on Oyster Production. American Journal of Agricultural Economics. 372-381. Web. 30 Oct. 2013.

Boudry, P., Collet, B., Cornette, F., Hervouet, V., Bonhomme, F., (2002) High variance in reproductive success of the Pacific oyster (Crassostrea gigas, Thunberg) revealed by microsatellite - based parentage analysis of multifactorial crosses. Aquaculture 204, 283-296.
Brazeau, D. A., and H. R. Lasker. (1992) Reproductive success in the Caribbean octocoral Briareum asbestinum. Mar. Biol. 114: 157-163.

Breitburg,D. L., M. A. Palmer,andT. Loher. (1995) Larval distribution sand the spatial patterns of settlement of an oyster reef fish: response to flow and structure.Marine Ecology Progress Series 125:45-60.

Bulleri F,  Bruno JF,  Benedetti-Cecchi L  (2008) Beyond Competition: Incorporating Positive Interactions between Species to Predict Ecosystem Invasibility. Plos Biol 6(6): e162.

Buroker, N. E. (1983). Sexuality with respect to shell length and group size in the japanese oyster crassostrea gigas. Malacologia, 23(2), 271-279. 
Butman,C. A., M. Fr6chette, W.R. Geyer,and V.R. Starczak. (1994) Flume experiments on food supply to the blue mussel Mytilus edulis L. as a function of boundary-layer flow. Limnology and Oceanography39:1755-1768.

Chavez, C. C., and Kang, Y. (2014). A simple epidemiological model. Discrete and Contonous System Series B, 19(1), 89-130.

Chia, F. -S., and L . R. Bickell.1 (1983). Echinodermata P.p . 545-620 in Reproductive Biology of Invertebrates, Volume II: Spermatogenesis and Sperm Function, K. G. Adiyodi and R. G. Adiyodi, eds. John Wiley and Sons, NY

Coen, L.D. and M.W. Luckenbach. (2000). Developing success criteria and goals for evaluating shellfish habitat restoration: Ecological function or resource exploitation? Ecol. Eng. 15: 323-343.

Courchamp, F., T. Clutton-Brock, and B. Grenfell. (1999). Inverse density dependence and the Allee effect. Trends Ecol. Evol. 14:405-410.

Cornell University. (n.d.). Dynamic Stability. Lecture. Retrieved from https://courses.cit.cornell.edu/mae5070/DynamicStability.pdf

Dartmouth College. (2012). DIFFUSION. Course Notes. Retrieved from http://engineering.dartmouth.edu/~d30345d/courses/engs43/chapter2.pdf

Csanady, G. T. 1973. Turbulent Diffusion in the Environment. D. Reidel, Dordrecht, Netherlands.

Davis, H. C., and Loosandoff , V. L. (1952). Temperature requirements for maturation of gonads of northern oysters. Bio Bullentin, 103, 80-96.

Denny, M. W. (1988). Biology and the Mechanics of the Wave-swept
Environment. Princeton University Press, N J. 329 pp.

Denny, M. W., and M. F. Shibata. (1989) Consequences of surf-zone turbulence for settlement and external fertilization. Am. Nat. 134: 859-889.

Diaz, R.J., and R. Rosenberg. (1995) Marine benthic hypoxia: a review of its ecological effects and the behavioral responses of benthic macrofauna. Oceanography and Marine Biology: An Annual Review 33:245-303.

Eckman, J.E. (1983) Hydrodynamic processes affecting benthic recruitment. Limnology and Oceanography 28:241-257.

Frankenberg, D. (1995)  Report of North Carolina Blue Ribbon Advisory Council on Oysters. North Carolina Department of Environmental Health and Natural Resources, Raleigh, North Carolina, USA.

Gaines, S., and J. Roughgarden. (1985) Larval settlement rate: a leading determinant of structure in an ecological community of the marine intertidal zone. Proc. Natl. Acad. Sci. USA 82:3707-3711.

Garvie, M. R. (2007). Finite-Difference Schemes for Reaction–Diffusion Equations Modeling Predator–Prey Interactions in MATLAB. Bulletin of Mathematical Biology,69(3), 931-956. doi:10.1007/s11538-006-9062-3

Gereffi , G., Lowe, M., Stokes, S., and Wunderink, S. (2012) Restoring Gulf Oyster Reefs. Duke University 1-60

Gosling, E., (2003), Bivalve Molluscs, Fishing News Books.

Grabowski, Johnathan H. (2004). “Habitat Complexity Disrupts Predator-Prey Interactions But Not the Trophic Cascade on Oyster Reefs." Ecology.: 995-1004.

Grabowski, J. H., and Peterson, H. C. (2007). “Restoring oyster reefs to recover ecosystem services.” Ecosystem Engineers.: 281-298

Grosberg, R . K. (1991) Sperm-mediated gene flow and the genetic structure of a population of the colonial ascidian Botryllus schlosseri. Evolution 45: 130-142.

Hargis, W. J., and D. S. Haven. (1988) The imperiled oyster industry of Virginia: a critical analysis with recommendations for restoration. Special report number 290 in applied marine science and ocean engineering. Virginia Sea Grant Marine Advisory Services. Virginia Institute of Marine Science, Gloucester Point, Virginia, USA.

Henderson, Jim and Jean O’Neil. (2003). Economic Values Associated with Construction of Oyster Reefs by the Corps of Engineers. Vicksburg, MS: U.S. Army Engineer Research and Development Center. \url{http://www.wes.army.mil/el/emrrp}.

Horton, T. and W.M. Eichbaum. (1991) Turning the Tide, Saving the Chesapeake Bay. Island Press, Washington, DC. 327 pp.

Huffaker,C. B. (1958) Experimental studies on predation :dispersion factors and predator-prey oscillations. Hilgardia 27:343-383.

Jackson, J.B.C., M.X. Kirby, W.H. Berger, K.A. Bjorndal, L.W. Botsford, B.J. Bourque, R.H. Bradbury, R. Cooke, J. Erlandson, J.A. Estes, T.P. Hughes, S. Kidwell, C.B. Lange, H.S. Lenihan, J.M. Pandolfi, C.H. Peterson, R.S. Steneck, M.J. Tegner, and R.R. Warner. (2001) Historical overfishing and the recent collapse of coastal ecosystems. Science 293:629-637.
Jones, C. G., J.H. Lawton,and M. Shachak. (1994) Organisms as ecosystemengineers. Oikos 69:373-386.

Kennedy, V S. Veliger   (1983) Sex ratios in oysters, emphasizing Crassostrea virginica from Chesapeake Bay, Maryland  329-338. Vol. 25, Iss. 4,

Kinne, 0.  (1963) The effect of temperature and salinity on marine and brackish water animals. Oceanography and Marine Biology: An Annual Review 1:301-340.

Lafferty, K. D., and Ward, J. R., (2004). The elusive baseline of marine disease: Are diseases in ocean ecosystem increasing. Plos Biol 2(4), 0542.

Lauzon-Guay, J., and Scheibling, R. E. (2007). Importance of Spatial Population Characteristics on the Fertilization Rates of Sea Urchins. The Biological Bulletin,212(3), 195-205. doi:10.2307/25066602

Levitan, D. R., M. A. Sewell, and F.-S. Chia. (1991) Kinetics of fertilization in the sea urchin Strongylocentrotus franciscanus: interaction of gamete dilution, age, and contact time. BiologicalBulletin 181:371-378.

Levitan, D.R., M.A. Sewell and F.-S. Chia. (1992) How distribution and abundance influence fertilization success in the sea urchin Strongylocentrotus franciscanus. Ecology 73(1):248-254.

Lenihan, H. S.. (1999). Physical-Biological Coupling on Oyster Reefs: How Habitat Structure Influences Individual Performance. Ecological Monographs, 69(3), 251–275.

Lenihan HS, Peterson CH. (1998). How habitat degradation through fishery
disturbance enhances impacts of hypoxia on oyster reefs. Ecological
Applications 8: 128–140.

Lenihan, H.S., C.H. Peterson, and J.M. Allen. (1996) Does flow speed also have a direct effect on the growth of active suspension feeders: an experimental test on oysters. Limnol. Oceanogr. 41:1359-1366.

Leonard, G. H., J. M. Levine, P. R. Schmidt,and M. D. Bertness. (1998) Flow-driven variation in intertidal community structure in a Maine estuary. Ecology 79:1395-1491.

Luckenbach Mark W., Coen Loren D., Ross P.G., and Stephen Jessica A. (2005). Oyster Reef Habitat Restoration: Relationships Between Oyster Abundance and Community Development based on Two Studies in Virginia and South Carolina. Journal of Coastal Research, 40, 64–78.

MacKenzie, C.L., Jr. (1996) Management of natural populations. Pp. 707-721 in V.S. Kennedy, R.I.E. Newell and A.F. Eble (eds.). The Eastern Oyster: Crassostrea virginica. Maryland Sea Grant College, College Park, MD.

Mann,K. H., andJ.R. M. Lazier. (1991) Dynamics of marine ecosystems: biological-physical interactions in the oceans. Blackwell, Boston, Massachusetts, USA.

Mann, R. and E. N. Powell. (2007) Why oyster restoration goals in the Chesapeake Bay are not and probably cannot be achieved. Journal of Shellfish Research 26:905- 917.

Marshall, D.J. (2002)  In situ measures of spawning synchrony and fertilization success in an intertidal, free-spawning invertebrate. Mar. Ecol. Prog. Ser. 236:113-119.

Mann, R., and Luckenbach, M. W. (2013). Sperm Swimming Speeds in the Eastern Oyster Crassostrea virginica(Gmelin, 1791). Journal of Shellfish Research,32(2), 387-390. doi:10.2983/035.032.0218

Matchar, E. (2018, January 10). As Storms Get Bigger, Oyster Reefs Can Help Protect Shorelines. 

Metaxas, A., R.E. Scheibling, and C.M. Young. (2002). Estimating fertilization success in marine benthic invertebrates: a case study with the tropical sea star Oreaster reticulates. Mar. Ecol. Prog. Ser. 226:87-101.

Meidel S. K (1999) Reproductive ecology of the sea urchin Strongylocentrotus drobachiensis. PhD. Dissertation, Dalhouseie University, Halifax California

Meidel, S. K., and R. E. Scheibling. (2001) Variation in egg spawning among subpopulations of sea urchins Strongylocentrotus droebachiensis: a theoretical approach. Mar. Ecol. Prog. Ser. 213: 97-110.

Meyer, D. L., Townsend, E. C. and Thayer, G. W. (1997), Stabilization and Erosion Control Value of Oyster Cultch for Intertidal Marsh. Restoration Ecology, 5: 93-99. doi:10.1046/j.1526-100X.1997.09710.x
Muschenheim, D. K. (1987) The dynamics of near-bed seston flux and suspension-feeding benthos. Journalof Marine Research45:473-496.

Newell, R. I. E., T. R. Fisher, R. R. Holyoke, and J. C. Cornwell. (2005)  Influence of eastern oysters on nitrogen and phosphorus regeneration in Chesapeake Bay, USA. pp. 93-120, in R. Dame and S. Olenin, editors. The Comparative Roles of Suspension Feeders in Ecosystems. Springer, Netherlands

Paynter, K.T. (1996). The effects of Perkinsus marinus infection on physiological processes in the eastern oyster, Crassostrea virginica. J. Shellfish Res. 15:119- 125.

Partnership for the Delaware Estuary (PDE). 2011. ‘Marine Bivalve Shellfish Conservation Priorities for the Delaware Estuary’. D. Kreeger, P. Cole, D. Bushek, J. Kraueter, J. Adkins. PDE Report \#11-03. 54 pp

Pavlos, N. V. (2004). Fertilization success in the eastern oyster (Crassostrea virginica) and hydrodynamic influences of oyster shell on larval retention(Unpublished master's thesis).

Pennington J, . T. (1985) The ecology of fertilization of echinoid eggs:
the consequence of sperm  dilution, adult aggregation and synchronous
spawning  Biol. Bull. 169: 417-430.

Jessica L. Petersen, Ana M. Ibarra, José L. Ramirez, Bernie May; An Induced Mass Spawn of the Hermaphroditic Lion-Paw Scallop, Nodipecten subnodosus: Genetic Assignment of Maternal and Paternal Parentage, Journal of Heredity, Volume 99, Issue 4, 1 July 2008, Pages 337–348

Powell, E.N., J.M. Morson, K.A. Aston-Alcox, and Y. Kim 2013. Accommodation of the sex-ratio in eastern oysters Crassostrea virginica to variation in growth and mortality across the estuarine salinity gradient. Journal of the Marine Biological Association of the United Kingdom, 93:533-555.

Ravit, B., Cooper, K., Buckley, B., Comi, M., and Mccandlish, E. (2014). Improving management support tools for reintroducing bivalve species (Eastern oyster [Crassostrea virginicaGmelin]) in urban estuaries. Integrated Environmental Assessment and Management,10(4), 555-565. doi:10.1002/ieam.1553

Roberts, Philip and Webster, Donald. (2003). Turbulent diffusion. Environmental Fluid Mechanics: Theories and Applications. 

Rodney, W.S. and K.T. Paynter. (2003) Comparison of fish trophic interactions on restored and unrestored oyster reef in the Chesapeake Bay, MD, USA. Abstract in the Book of Abstracts from the 17th Biennial Conference of the Estuarine Research Federation entitled Estuaries on the Edge, Convergence of Ocean, Land and Culture. Port Republic, MD, p. 113.

Roughgarden, J., S. Gaines, and S. Pacala. (1987) Supply side ecology: the role of physical transport processes. Br. Ecol. Soc. Symp. 27:491-518

Sanford,E., D. Bermudez,M. D. Bertness,and S. D. Gaines. (1994) Flow, food supply, and acorn barnacle population dynamics. Marine Ecology Progress Series 104:49-62.

Sanjeeva Raj, P.J. Oysters in a new classification of keystone species. Reson 13, 648–654 (2008). https://doi.org/10.1007/s12045-008-0071-4

Sebastiano, D. (2012) Quantifying the nutrient bioextraction capacity of restored eastern oyster populations in two coastal bays on Long Island, New York. Ms thesis, Stony Brook, NY, USA.

Skilleter,G. A., and C. H. Peterson. (1994) Control of foraging behavior of individuals with in an ecosystem context: the clam Macoma balthica and interactions between competition and siphon cropping. Oecologia 100:268-278.

Stokes, S., Wunderink, S., Lowe, M., and Gereffi, G. (2012, June). RESTORING GULF OYSTER REEFS. Retrieved from 

Styles, R. (2015). Flow and Turbulence over an Oyster Reef. Journal of Coastal Research,314, 978-985. doi:10.2112/jcoastres-d-14-00115.1

Thayer,G. W., editor. (1992) Restoring the nation's marine environment. Maryland Sea Grant, College Park, Maryland, USA.

Underwood, A. J. ,and E. J. Denley. (1984) Paradigms, explanations,and generalizations about models for the structure of intertidal communities on rocky shores. Pages 151-180 in D. R. Strong, L. Abele, and D. Simberloff,  editors. Ecological communities: conceptual issues and the evidence. Princeton University Press, Princeton, New Jersey, USA.

Vogel,S. (1981) Life in moving fluids: The physical biology off low. Princeton University Press, Princeton, New Jersey, USA

Waldman, J.  (1999)   Heartbeats in the Muck.  The Lyons Press, NY.

Wildish,D. J.,and D. D. Kristmanson. (1979) Tidal energy and sublittoral macrobenthic animals in estuaries. Journal of the Fisheries Research Board of Canada 36:1197-1206.

Yozzo, D., P. Wilber, and R.J. Will. ( 2004)  Beneficial use of dredged material for habitat
creation , enhancement, and restoration in New York-New Jersey Harbor.  Journal of
Environmental Management 73:39-52.

Yund P . (1990) An in situ measurement of sperm dispersal in a colonial
marine hydroid  J.  Exp. Zool. 253: 102-106
%\end{multicols}
\end{document}